\documentclass[prd,twocolumn,superscriptaddress,amsfonts,amssymb,amsmath,showpacs]{revtex4-2}

\usepackage{bm}
\usepackage{latexsym}
\usepackage[latin1]{inputenc}
\usepackage{graphicx}
\usepackage{amsmath}
\usepackage{amssymb}
\usepackage{amsthm}
\usepackage{relsize}
\usepackage{palatino}
\usepackage{mathpazo}
\usepackage{textcomp}
\usepackage{float}
\usepackage{booktabs}
\usepackage{dcolumn}
\usepackage{booktabs}
\usepackage{multirow}
\usepackage{tikz}
\usepackage{hyperref}
\hypersetup{
    colorlinks=true,
    linkcolor=purple,
    filecolor=magenta,      
    citecolor=blue
}
\usepackage{amsmath}
\usepackage{xcolor}
\usepackage{orcidlink}
\usepackage[caption=false]{subfig}
\usepackage{commath}
\makeatletter
\long\def\dddddot#1{%
  {\mathop {#1}\limits ^{\vbox to-1.4\ex@ {\kern -\tw@ \ex@ \hbox {\normalfont .....}\vss }}}%
}
\long\def\multidots#1#2{%
  \count@=0
  {{\mathop {#2}\limits ^{\vbox to-1.4\ex@ {\kern -\tw@ \ex@ \hbox {\normalfont %
  \loop%
  \ifnum#1>\count@%
  .%
  \advance\count@ by1%
  \repeat%
  }\vss }}}}%
}
\makeatother

\begin{document}

\title{\bf Exploring the Viability of \texorpdfstring{$f(Q,T)$}{} Gravity: Constraining Parameters with Cosmological Observations}

%\title{\bf Beyond $\Lambda$CDM: Constraining and Examining $f(Q, T)$ Gravity Model through Cosmological Observations}

\author{Rahul Bhagat\orcidlink{0009-0001-9783-9317}}
\email{rahulbhagat0994@gmail.com}
\affiliation{Department of Mathematics, Birla Institute of Technology and Science, Pilani, Hyderabad Campus, Jawahar Nagar, Kapra Mandal, Medchal District, Telangana 500078, India.}
\author{Santosh V. Lohakare \orcidlink{0000-0001-5934-3428}}
\email{lohakaresv@gmail.com}
\affiliation{Department of Mathematics, Birla Institute of Technology and Science, Pilani, Hyderabad Campus, Jawahar Nagar, Kapra Mandal, Medchal District, Telangana 500078, India.}
\author{B. Mishra\orcidlink{0000-0001-5527-3565}}
\email{bivu@hyderabad.bits-pilani.ac.in}
\affiliation{Department of Mathematics, Birla Institute of Technology and Science, Pilani, Hyderabad Campus, Jawahar Nagar, Kapra Mandal, Medchal District, Telangana 500078, India.}

\begin{abstract} In this paper, we explore the model of $f(Q,T)$ gravity, an extension of symmetric teleparallel gravity where the nonmetricity scalar $Q$ is non-minimally coupled to the trace of the energy-momentum tensor $T$. To ensure general covariance and theoretical consistency, we adopt the covariant formulation of $f(Q, T)$ gravity, which allows a coordinate-independent treatment of the field equations and facilitates the correct identification of effective energy-momentum components. The model is developed as an alternative to the standard $\Lambda$CDM cosmological model and is analyzed using Cosmic Chronometer and Pantheon$^+$ supernovae datasets. Through Markov Chain Monte Carlo analysis, we constrain the model parameters $\alpha$, $\beta$, and $H_0$, and compare the performance of the model with $\Lambda$CDM by evaluating statistical measures such as chi-square, Akaike information criterion (AIC), and Bayesian information criterion (BIC). The results show that the $f(Q, T)$ model effectively mimics $\Lambda$CDM while offering an alternative explanation based on modified gravity. We also examine cosmographic parameters like the deceleration parameter, confirming the transition of the Universe from deceleration to acceleration, and the violation of the strong energy condition, which aligns with observed late-time cosmic acceleration. Additionally, the model provides age estimates for the Universe that are consistent with current observations.

\end{abstract}

\maketitle

\section{Introduction}
The discovery in the late 20th century of the accelerated expansion \cite{Riess_1998_116, Perlmutter_1998_517} of the Universe is one of the most surprising facts in modern cosmology. This finding defied expectations because gravity, which is always attractive, ought to be working against the expansion of the Universe and at some stage it would bring it to a complete stop. In that situation, gravitational attraction would take over, and there would be a contraction stage. In contrast, several different cosmological evidences, such as Type Ia supernovae \cite{Ade_2016_594a, Abbott_2016_460}, cosmic microwave background \cite{Larson_2011_192, Komatsu_2011_192}, large-scale structure surveys \cite{Eisenstein_2005_633, Hinshaw_2013_208} have convincingly shown that in recent cosmological time, the Universe has entered into a period of accelerated expansion. This is a surprising behavior of the Universe, which suggests gaps in our current knowledge of the physical laws, and new theories should be developed to make them consistent.

Several theories have been developed in the last twenty years to explain the accelerated expansion \cite{Spergel_2003_148, Tegmark_2004_69, Hinshaw_2013_208, Lohakare_2025_ApJ, Narawade:2023pze} and bouncing scenario \cite{Agrawal_2023_83, Agrawal:2021_bouncing}. At a very high level, these activities can be divided into two groups. The initial approach aims to clarify the matter-energy structure of the Universe, positing the presence of an elusive component often referred to as dark energy \cite{Abbott_2018_480}. This concept seeks to account for the observed accelerated expansion of the Universe, which cannot be explained solely by the known forms of matter and energy. This part is believed to have a negative pressure and provides a repulsive form of gravitational energy, enabling not just a continued expansion, but an acceleration of the process. A common explanation within this framework involves introducing the cosmological constant ($\Lambda$) \cite{Fujii_1982_26} into the Einstein field equations of General Relativity. The literature on modified gravity encompasses a diverse range of theoretical frameworks. Among these, a prominent class of theories is formulated within the symmetric teleparallel formalism \cite{Nester_1999}. Unlike approaches that rely on spacetime curvature \cite{Sotiriou_2008_82} or torsion \cite{Cai_2016_79}, this formalism employs nonmetricity as the geometric entity governing gravitational interactions. In particular, the $f(Q)$ gravity theory modifies the gravitational action by introducing a general function of the nonmetricity scalar $Q$ \cite{Jimenez_2018_98}, which quantifies how distances change under parallel transport. This formulation allows for a self-consistent explanation of cosmic acceleration without requiring the dark energy component. 

A further extension of this approach is found in $f(Q, T)$ gravity \cite{Xu_2019_79}, where the gravitational action depends not only on the nonmetricity scalar $Q$ but also on the trace of the energy-momentum tensor $T$. This additional coupling between gravity and the energy-momentum content of the Universe allows for a more comprehensive exploration of the interaction between matter and spacetime geometry. The $f(Q)$ and $f(Q,T)$ gravity models represent promising avenues for addressing the fundamental issues of modern cosmology, including the cosmological constant problem and the nature of dark energy \cite{Dimakis_2021_38, Lin_2021_103, heisenberg_2023_review}, within a modified gravity framework. These theories offer new insights into the behavior of gravity and its role in the evolution of the Universe.

Recent investigations into $f(Q, T)$ gravity have highlighted its potential to address key cosmological phenomena. Shiravand et al. \cite{Shiravand_2024} explored insights into the early inflation of the Universe and examined late-time cosmic acceleration using a time-dependent deceleration parameter supported by observational data \cite{Khurana_2024}. Najera et al. \cite{N_jera_2022} investigated the implications of matter Lagrangian degeneracy and its effects on the Symmetric Teleparallel Equivalent of General Relativity, while Pati et al. \cite{Pati_2023_83} employed dynamical system analysis to study the accelerating behavior of cosmological parameters. Sharif et al. \cite{Sharif_2024} reconstructed the theory using ghost dark energy models to explore the role of dark energy in cosmic dynamics. Together, these studies and others reflect the growing interest in $f(Q, T)$ gravity as a versatile framework for addressing significant challenges in cosmology, including early inflation, late-time acceleration, and the enigmatic nature of dark energy \cite{Zia_2021, Bhagat_2023_42, Shiravand_2022_37, Bhagat_ASPdyna2024}.

This paper aims to study the accelerated expansion of the Universe using modified $f(Q, T)$ gravity as an alternative to the standard $\Lambda$CDM model. The structure of this paper is outlined as follows: Section \ref{Sec:2} provides an overview of the symmetric teleparallel equivalent of General Relativity and its extension to $f(Q, T) $ gravity. We also mention the fundamental equations governing FLRW cosmology within this framework. In Section \ref{Sec:3}, we present a detailed analysis of the $f(Q, T)$ gravity model, using observational data sets from cosmic chronometers (CC) \cite{Moresco_2022_25} and Pantheon$^+$ \cite{Brout_2022_938}. To assess the model performance, we compare it with the standard $\Lambda$CDM model. Additionally, we investigate key cosmological parameters, including the deceleration parameter and statefinder diagnostics, to establish the strength of $f(Q, T) $ as a compelling alternative gravity model. Further tests are conducted, examining the behavior of effective pressure, density parameters, and the violation of the strong energy condition (SEC). Finally, in Section \ref{Sec:4}, we present our conclusions, discussing the implications of our findings and emphasizing the model consistency with observational data.

\section{Background equations of \texorpdfstring{$f(Q, T)$}{} theory}\label{Sec:2}

In the framework of $f(Q,T) $ gravity \cite{Xu_2019_79,Loo2023}, the action is given by
\begin{eqnarray} \label{action}
S = \int \left[ \frac{1}{2\kappa}f(Q,T)\sqrt{-g}~d^4x + \mathcal{L}_m\sqrt{-g}~d^4x \right],
\end{eqnarray}
where $\mathcal{L}_m $ represents the matter Lagrangian and $g $ is the determinant of the metric tensor $g_{\mu\nu} $. $\kappa$ is the gravitational coupling constant.
Varying the action \eqref{action} with respect to the metric tensor leads to the generalized field equations of motion for $f(Q, T)$ gravity. The resulting metric field equation is

\begin{multline}\label{Eqn.2}
    \frac{2}{\sqrt{-g}} \, \partial_\lambda \left( \sqrt{-g} f_Q P^\lambda_{\mu\nu} \right) - \frac{1}{2} f \, g_{\mu\nu} + f_T (T_{\mu\nu} + \Theta_{\mu\nu}) \\+ f_Q \left( P_{\nu\rho\sigma} Q_\mu^{\rho\sigma} - 2 P^{\rho\sigma\mu} Q_{~\nu}^{\rho\sigma} \right) = \kappa T_{\mu\nu},
\end{multline}

where $f_Q = \partial f / \partial Q$ and $f_T = \partial f / \partial T$ represent partial derivatives of the Lagrangian. The superpotential, the non-metricity tensor, and the energy-momentum tensor are expressed as
\begin{eqnarray}\label{eq.3}
P^{\lambda}_{~\mu\nu} &=& -\frac{1}{2}L^{\lambda}_{~\mu \nu}+\frac{1}{4}\left(Q^{\lambda}-\tilde{Q}^{\lambda}\right)g_{\mu \nu}-\frac{1}{4}\delta^{\lambda}_{(\mu}Q_{\nu)},\nonumber\\
Q_{\lambda} &=& Q_{\lambda}^{\;\;\mu}\;_{\mu},~~~~~~~ \tilde{Q}_{\lambda}=Q^{\mu}\;_{\lambda\mu},\nonumber\\
T_{\mu \nu}&=&\frac{-2}{\sqrt{-g}} \frac{\delta(\sqrt{-g}L_{m})}{\delta g^{\mu \nu}},~~~~~~~\Theta_{\mu \nu}=g^{\lambda l}\frac{\delta T_{\lambda l}}{\delta g^{\mu \nu}}.
\end{eqnarray}

 The non-metricity scalar is defined as
 \begin{eqnarray}
    && Q = Q_{\lambda\mu\nu}P^{\lambda\mu\nu} \nonumber\\
    && =\frac{1}{4}(-Q^{\lambda\nu\rho}Q_{\lambda\nu\rho}+2Q^{\lambda\nu\rho}Q_{\rho\lambda\nu}-2Q^\rho\tilde{Q_\rho}+Q^\rho Q_\rho)
\end{eqnarray}

Varying the gravitational action \eqref{action} with respect to the connection yields the field equations of $f(Q, T) $ gravity

\begin{equation}
    \nabla_\mu \nabla_\nu \left( 2\sqrt{-g} \, f_Q \, P^{\mu\nu}_\lambda - \kappa \, H_\lambda^{\mu\nu} \right) = 0,
\end{equation}

where $H^\lambda_{\mu\nu}$ represents the hypermomentum tensor density, defined as

\begin{equation}
   H_\lambda^{\mu\nu} = \frac{\sqrt{-g}}{2\kappa} f_T \frac{\delta T}{\delta \Gamma^\lambda_{\mu\nu}} + \frac{\delta(\sqrt{-g} \mathcal{L}_M)}{\delta \Gamma^\lambda_{\mu\nu}}. 
\end{equation}

\subsection{Covariant formulation of $f (Q, T)$}
The covariant formulation of $f(Q, T)$ gravity is crucial as it ensures that the field equations hold true across all coordinate systems, preserving the foundational principle of general covariance from general relativity. Unlike the coincident gauge, which simplifies calculations by setting the connection to zero but restricts applicability to specific spacetimes like FLRW, the covariant approach allows for a consistent and flexible treatment of more general geometries. It also enables a proper formulation of effective energy density and pressure, while accurately capturing the energy-momentum conservation law, which earlier treatments had formulated incompletely. Recognizing these advantages, in this paper, we adopt the covariant formulation given in \cite{Loo2023} to build a more robust and comprehensive framework for exploring the cosmological implications of $f(Q, T)$ gravity.

The fundamental assumption of the geometry is: the spacetime is curvature-free and torsion-free. These conditions lead to the following constraints

\begin{equation}\label{Eq.6}
 \mathring{R}_{\mu\nu} + \overset{\circ}{\nabla}_\alpha L^\alpha_{\mu\nu} - \overset{\circ}{\nabla}_\nu \tilde{L}_\mu + \tilde{L}^\alpha L^\alpha_{\mu\nu} - L_{\alpha\beta\nu} L^{\beta\alpha}_{\mu} = 0,
\end{equation}

\begin{equation}\label{Eq.7}
    \mathring{R} + \overset{\circ}{\nabla}_\alpha (L^\alpha - \tilde{L}^\alpha) - Q = 0,
\end{equation}
where $\mathring{R}_{\mu\nu}$ and $\mathring{R}$ denote the Ricci tensor and scalar derived from the Levi-Civita connection. Under the coincident gauge, the affine connection is set to zero ($\Gamma^\lambda_{\mu\nu} = 0$), implying that the Levi-Civita connection is entirely given by minus the disformation tensor, $\mathring{\Gamma}^\lambda_{\mu\nu} = -L^\lambda_{\mu\nu}$. This simplifies the divergence of the determinant of the metric: $\partial_\lambda \sqrt{-g} = -\sqrt{-g} \tilde{L}_\lambda.$

Using these results, one can evaluate the divergence term involving the superpotential tensor and the derivative of the Lagrangian
\begin{multline}
    \frac{2}{\sqrt{-g}} \, \partial_\lambda \left( \sqrt{-g} f_Q P^\lambda_{\mu\nu} \right) + f_Q \left( P_{\nu\rho\sigma} Q_\mu^{\rho\sigma} - 2 P^{\rho\sigma\mu} Q_{~\nu}^{\rho\sigma} \right)\\= 2(\nabla_\lambda f_Q) P^\lambda_{\mu\nu} + 2 f_Q (\overset{\circ}{\nabla}_\lambda P^\lambda_{\mu\nu} - \tilde{L}_\alpha L^\alpha_{\nu\mu} + L_{\alpha\beta\nu} L_{~\mu}^{\beta\alpha}).
\end{multline}

Separately using the superpotential tensor and Eqn. (\ref{Eq.6}) and (\ref{Eq.7}), a direct computation from the definitions yields
\begin{equation}
    2\overset{\circ}{\nabla}_\alpha P^\alpha_{\mu\nu} = \mathring{R}_{\mu\nu} + \frac{Q - \mathring{R}}{2}g_{\mu\nu} + \tilde{L}_\alpha L^\alpha_{\nu\mu} + L_{\alpha\beta\nu} L_{~\mu}^{\beta\alpha}.
\end{equation}

By combining the above two expressions, we arrive at a fully covariant form
\begin{multline}
  \frac{2}{\sqrt{-g}} \partial_\lambda (\sqrt{-g} f_Q P^\lambda_{\mu\nu}) +  f_Q \left( P_{\nu\rho\sigma} Q_\mu^{\rho\sigma} - 2 P^{\rho\sigma\mu} Q_{~\nu}^{\rho\sigma} \right)\\
= 2(\nabla_\lambda f_Q) P^\lambda_{\mu\nu} + f_Q \left( \mathring{R}_{\mu\nu} - \frac{\mathring{R} - Q}{2} g_{\mu\nu} \right). 
\end{multline}

This ultimately allows us to express the field equations of $f(Q, T)$ gravity in metric field equation form
\begin{multline}\label{Eq.11}
   f_Q \mathring{G}_{\mu\nu} + \frac{1}{2} g_{\mu\nu}(Q f_Q - f) + f_T(T_{\mu\nu} + \Theta_{\mu\nu})\\ + 2(\nabla_\lambda f_Q) P^\lambda_{\mu\nu} = \kappa T_{\mu\nu}, 
\end{multline}
where $\mathring{G}_{\mu\nu}$ is the Einstein tensor derived from the Levi-Civita connection, and $\Theta_{\mu\nu}$ is related to the variation of the matter energy-momentum tensor. From this, we define the effective stress-energy tensor in a form that cleanly encapsulates all additional contributions from the matter-geometry coupling
\begin{multline}\label{Eq.12}
    \kappa T^{\text{eff}}_{\mu\nu} = \kappa T_{\mu\nu} - f_T (T_{\mu\nu} + \Theta_{\mu\nu}) - \frac{1}{2} g_{\mu\nu}(Q f_Q - f) \\- 2(f_{QQ} \nabla_\lambda Q + f_{QT} \nabla_\lambda T) P^\lambda_{\mu\nu}.
\end{multline}

This formulation not only preserves the general covariance of the theory but also allows us to clearly interpret gravitational dynamics and matter interactions in terms of effective energy density and pressure.

We consider the Friedmann-Lema\'itre-Robertson-Walker (FLRW) spacetime
\begin{eqnarray}\label{Eq.13}
ds^2 = -dt^2 + a^2(t)(dx^2 + dy^2 + dz^2),
\end{eqnarray}
where $a(t) $ is the scale factor, and $H = \frac{\dot{a}}{a} $ is the Hubble parameter. For the flat FLRW metric, the nonmetricity scalar $Q = -6H^2 $. The most commonly used matter component is the perfect cosmic fluid whose energy-momentum tensor is,
\begin{equation}
T_{\mu\nu}=(\rho+p)u_{\mu}u_{\nu}+pg_{\mu\nu}.
\end{equation}
Here, $\rho $ and $p$\ respectively represent the energy density and isotropic pressure of the perfect, and $u^{\mu }=(1,0,0,0)$ represents the four-velocity vector components characterizing the fluid and $\Theta_{\mu\nu}=pg_{\mu\nu}-2T_{\mu\nu}$. By applying the covariant field Eq.(\ref{Eq.11}) to the spatially flat FLRW metric from Eq.(\ref{Eq.13}), we obtain the modified Friedmann equations that govern the cosmological dynamics in $f(Q, T)$ gravity. These take the form
\begin{equation}\label{Eq.14}
    (\kappa + f_T)\rho + f_T p = \frac{f}{2} + 6H^2 f_Q,
\end{equation}
\begin{equation}\label{Eq.15}
    \kappa p = -\frac{f}{2} - 6H^2 f_Q - \frac{d}{dt}(2H f_Q),
\end{equation}

These equations encapsulate the influence of both geometric corrections and matter coupling on cosmic expansion. Furthermore, by using the effective energy-momentum tensor defined in Eq.(\ref{Eq.12}), one can express the dynamics in terms of effective energy density and pressure, which absorb all the additional contributions arising from the modified gravity terms. These are given by
\begin{equation}
    \kappa \rho_{\text{eff}} = (\kappa + f_T)\rho + f_T p - \frac{f}{2} - 3H^2 f_Q,
\end{equation}
\begin{equation}
    \kappa p_{\text{eff}} = \kappa p + \frac{f}{2} + 3H^2 f_Q + 2H \dot{f}_Q.
\end{equation}

This effective formulation allows for a more intuitive interpretation of the gravitational modifications as corrections to the standard cosmological fluid properties, making it easier to analyze the evolution of the universe within this extended theoretical framework.

\section{Observational Analysis of the \texorpdfstring{$f(Q, T)$}{} Model}\label{Sec:3}
Assuming that the Universe contains only dust-like matter, we apply the exponential form of the functional $f(Q, T)$ as $f(Q, T) = Q~e^{\alpha \frac{Q_0}{Q}}+\beta T$ \cite{Xu_2020_80,BHAGAT2025101913,Bohmer_2023_9} in the Friedmann Eqs. (\ref{Eq.14}) and (\ref{Eq.15}). With this consideration $\kappa=1$, we derive the expression for the Hubble parameter that provides a first-order ordinary differential equation. This equation describes the expansion of the Universe as a function of time. We transform it into redshift to make it more convenient for comparison with observational data. For this transformation, we use the relation $\frac{d}{dt} = -(1+z)H \frac{d}{dz}$. Hence,
\begin{equation}\label{eq:16}
    H\;'(z)=\frac{3H_0(1+\beta)}{(z+1)(2+3\beta)} \times \frac{E(z)^3 \left(E(z)^2-2 \alpha \right)}{(2 \alpha ^2-\alpha  E(z)^2+E(z)^4)}
\end{equation}
where  $(\,'\,)$ denotes differentiation with respect to redshift $z$,  $H_0$ is the present value of the expansion rate of the Universe, $E(z)=\frac{H(z)}{H_0}$ and $\alpha$ and $\beta$ is the model parameter of the $f(Q, T)$ model. Also,
\begin{equation}\label{eq:17}
\rho(z)=\frac{6~{H_0}^2 e^{{\alpha }/{E(z)^2}} \left(E(z)^2-2 \alpha  \right)}{3 \beta +2}
\end{equation}

To solve the differential equation, we employ the Odeint method for numerical integration. Furthermore, we perform Markov Chain Monte Carlo (MCMC) analysis to constrain the model parameters of the $f(Q, T) $ model, including the initial conditions of the differential equation. MCMC allows us to explore the parameter space efficiently, estimate the best-fit values, and quantify uncertainties by sampling from the posterior distribution using observational data. This enables a robust comparison with cosmological data sets, ensuring that the model is consistent with the data.

\subsection{MCMC Analysis}
MCMC analysis is a powerful statistical method used to estimate the probability distributions of model parameters by generating a chain of samples from a target posterior distribution. The foundation of MCMC lies in Bayesian inference, where the goal is to obtain the posterior probability distribution of model parameters given some data. This posterior distribution is proportional to the likelihood of the data given the parameters, multiplied by the prior distribution of the parameters
\begin{eqnarray}
P(\theta | \text{data}) \propto P(\text{data} | \theta) \cdot P(\theta) \, ,
\end{eqnarray}
where $\theta $ represents the model parameters, $P(\theta | \text{data}) $is the posterior, $P(\text{data} | \theta) $ is the likelihood, and $P(\theta) $ is the prior distribution. The posterior outlines both prior knowledge about the parameters and the information provided by the observational data.

In this analysis, we aim to constrain the cosmological parameters $\alpha $, $\beta $, and $H_0 $ using two key data sets: the CC data and the Pantheon$^+$ supernova data. The CC data set provides 32 data points for the Hubble parameter, derived from the age difference between passively evolving galaxies at different redshifts, up to $z \approx 2$ \cite{Jimenez_2002_573}. This method allows for direct measurements of the Hubble parameter without relying on the Cepheid distance scale or specific cosmological models, although it does depend on stellar population synthesis models. Meanwhile, the Pantheon$^+$ data set consists of 1701 Type Ia supernovae \cite{Brout_2022_938}, with redshift values ranging from $0.01 $ to $2.3 $. The observed supernova magnitudes are corrected for systematics such as light-curve stretch, color, and host galaxy mass. By combining the CC and Pantheon$^+$ data sets, we seek to robustly estimate the parameters $\alpha $, $\beta $, and $H_0 $ using a comprehensive cosmological model.

To assess the fit of the theoretical model to the observed data, we compute the $\chi^2$ statistic for each data set. For the CC data set, the $\chi^2$ function is given by
\begin{eqnarray}
\chi^2_H(\mathbf{p}) = \sum_{i=1}^{32} \frac{[H(z_i, \mathbf{p}) - H_{\text{obs}}(z_i)]^2}{\sigma_H^2(z_i)} \, ,
\end{eqnarray}
where $H(z_i, \mathbf{p}) $ represents the theoretical Hubble parameter at redshift $z_i $ for the parameter set $\mathbf{p} = (\alpha, \beta, H_0) $, $H_{\text{obs}}(z_i) $ is the observed Hubble parameter at redshift $ z_i $, and $ \sigma_H(z_i) $ is the associated uncertainty in the observation. This equation quantifies the square deviations between the theoretical and observed values, weighted by the observational uncertainties.
 \begin{figure}
    \centering
    \includegraphics[width=8cm]{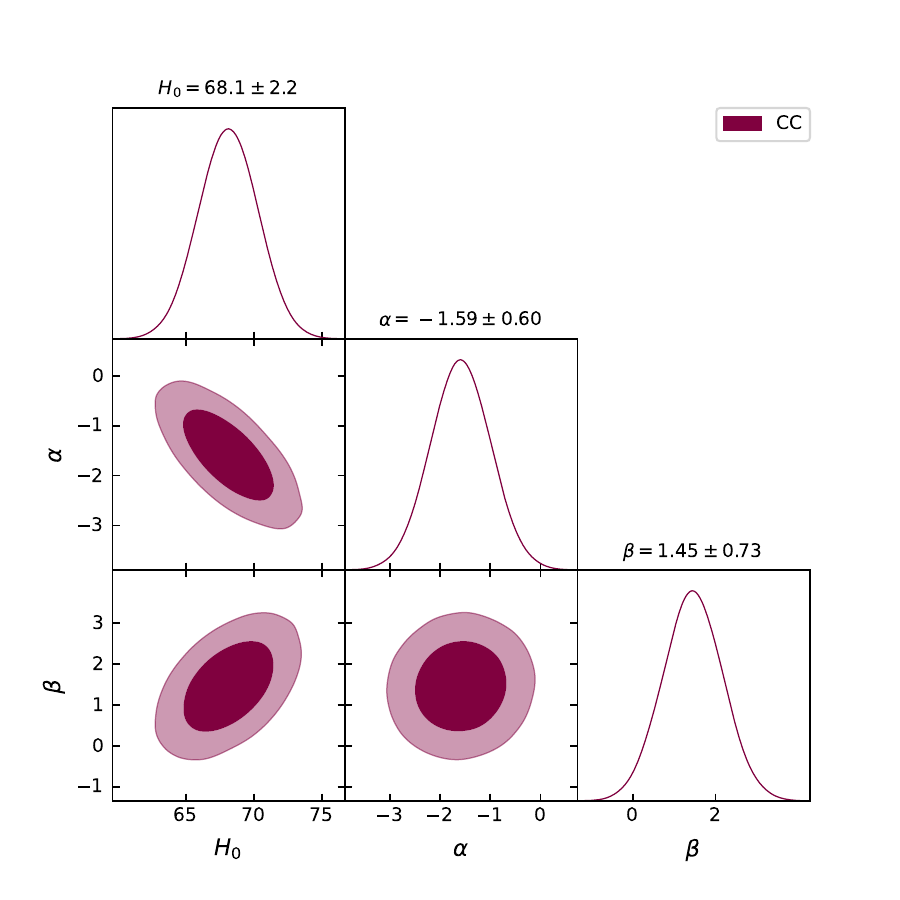}
    \includegraphics[width=8cm]{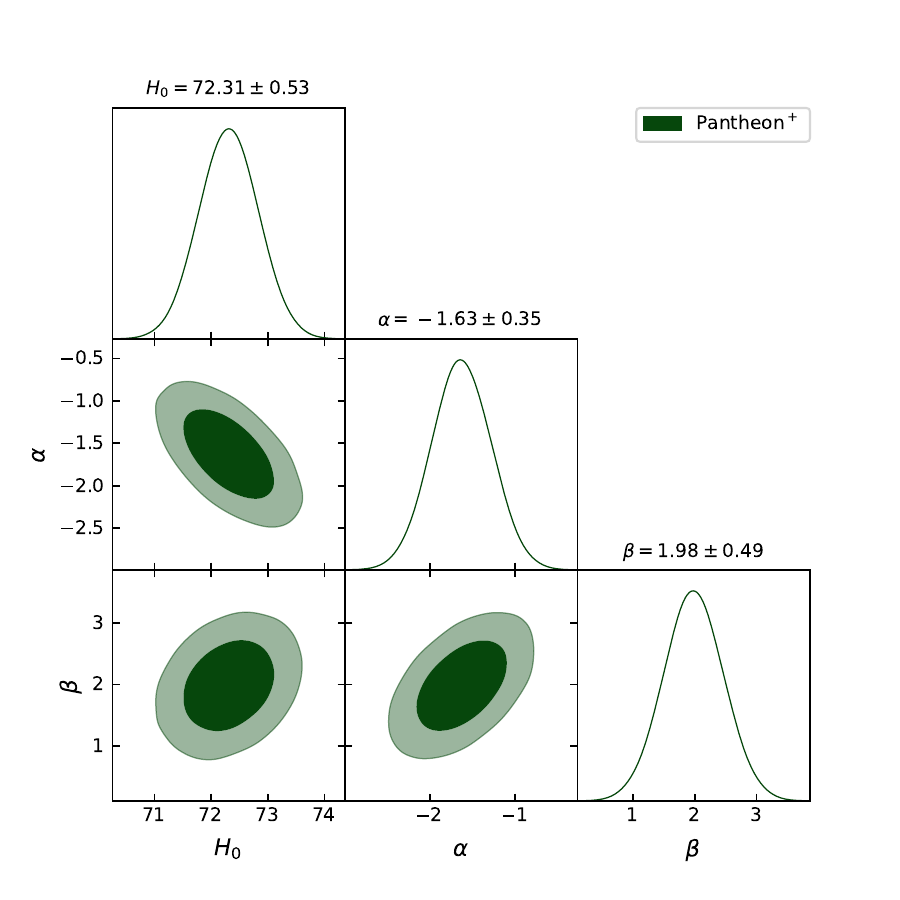}
    \includegraphics[width=8cm]{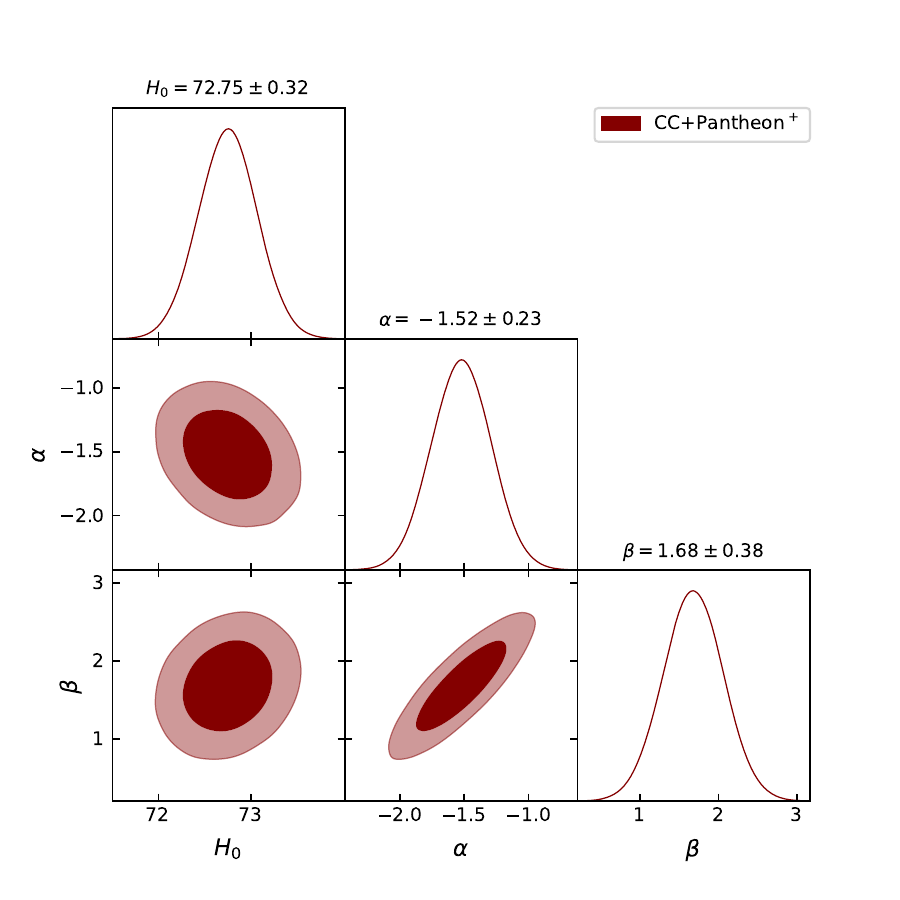}
   {\\caption{Contour plot for CC, Pantheon$^+$ and combine of CC$+$Pantheon$^+$ data set.}}
    \label{Fig1} 
 \end{figure}

For the Pantheon$^+$ supernova dataset, the $\chi^2$ function is formulated to incorporate the distance modulus $ \mu(z) $, which is determined by the luminosity distance. The $\chi^2$ function is defined as
\begin{multline}
  \chi^2_{\text{Pantheon$^+$}}(\mathbf{p}) = \Delta \mu(z, \mathbf{p})^T C_{\text{Pantheon$^+$}}^{-1} \Delta \mu(z, \mathbf{p}) \\+ \ln \left( \frac{S}{2\pi} \right) - \frac{k(\mathbf{p})^2}{S}  \, ,
\end{multline}
where $ \Delta \mu(z, \mathbf{p}) = \mu(z, \mathbf{p}) - \mu_{\text{obs}}(z) $, and $ \mu(z, \mathbf{p}) = 5 \log_{10} \left[ D_L(z, \mathbf{p}) \right] + M $ is the theoretical distance modulus. The absolute magnitude $ M $ is treated as a nuisance parameter in the analysis. The luminosity distance $ D_L(z, \mathbf{p}) $ is given by
\begin{equation}
D_L(z, \mathbf{p}) = c (1 + z) \int_0^z \frac{dz'}{H(z', \mathbf{p})} \, ,
\end{equation}
where $ c $ is the speed of light, and $ H(z', \mathbf{p}) $ is the theoretical Hubble parameter, $ C_{\text{Pantheon$^+$}} $ is the covariance matrix of the Pantheon$^+$ data, $ S $ is the sum of all elements in $ C_{\text{Pantheon$^+$}}^{-1} $, and $ k(\mathbf{p}) $ is defined as
\begin{eqnarray}
k(\mathbf{p}) = \Delta \mu(z, \mathbf{p})^T \cdot C_{\text{Pantheon$^+$}}^{-1} \, .
\end{eqnarray}

By computing these $\chi^2$ values for both the CC and Pantheon$^+$ data sets, we can compare how well the theoretical model fits each data set individually and in combination.

To further evaluate the model and account for its complexity, we compute the Akaike information criterion (AIC) \cite{Akaike_1974} and Bayesian information criterion (BIC) \cite{Schwarz_1978}. These criteria provide a means of balancing the goodness of fit with the number of free parameters in the model. The AIC is calculated as
\begin{eqnarray}
\text{AIC} = \chi^2_{\text{min}} + 2k \, ,
\end{eqnarray}
where $ \chi^2_{\text{min}} $ is the minimum chi-square value and $ k $ is the number of free parameters. The BIC is defined similarly, with a term that depends on the number of data points, $N$. It is given by 
\begin{eqnarray}
\text{BIC} = \chi^2_{\text{min}} + k \ln(N) \, ,
\end{eqnarray}
where $N$ represents the number of data points (32 for CC and 1701 for Pantheon$^+$), and $k$ is the number of free parameters. Lower values of AIC and BIC indicate a better model fit, with BIC imposing a stricter penalty for the number of free parameters compared to AIC. These criteria improve the evaluation of the fit of the model to the data and the selection of the most efficient model that accurately represents the observations.
\begin{table*}[htb]
\renewcommand\arraystretch{1.5}
\centering % used for centering table

\begin{tabular}{|c|c|c|c|c|c|c|c|c|c|c|} % centered columns (3 columns)
\hline %inserts double horizontal lines
~~~Data sets~~~& ~~~$H_0$  ~~~& ~~~~$\alpha$~~~~ & ~~~$\beta$~~~ &~~~Age Of the Universe~~~ &$\chi^2_{\text{min}}$&AIC& BIC&$\triangle\chi^2_{\text{min}}$&$\triangle $AIC&$\triangle$BIC
     \\ [0.5ex] % inserts table
%heading
% inserts single horizontal line
\hline\hline
CC & $68.1 \pm 2.2$ & $-1.59 \pm 0.60$ &  $1.45\pm 0.73$&$13.831~~Gyr$&$16.51$& $22.51$ &$26.81$ & $1.97$&$3.97$& $5.4$\\
\hline
Pantheon$^+$ & $72.31\pm0.53$ &  $-1.63\pm0.35$ & $1.98 \pm 0.49$& $13.898~~Gyr$&$816.53$&$822.53$&$838.85$ &$3.92$& $5.92$& $11.36$ \\
\hline
CC $+$ Pantheon$^+$ & $72.75\pm0.32$ &  $-1.52\pm0.23$ & $1.68\pm0.38$& $13.851~~Gyr$&$847.81$&$859.81$&$880.12$ &$1.18$& $5.18$& $11.75$ \\
% \hline
% Priors & $(60,80)$ &  $(-10,10)$ & $(-10,10)$& $-$&$-$&$-$&$-$&$-$&$-$\\
\hline %inserts single line
\end{tabular}
\caption{Constrained parameter values, and $\chi^2_{\text{min}}$ for CC and pantheon$^+$ data sets and its difference with $\Lambda$CDM.} 
\label{tableA1}
\end{table*}

\begin{figure}[H]
\centering
\includegraphics[width=8.5cm,height=5cm]{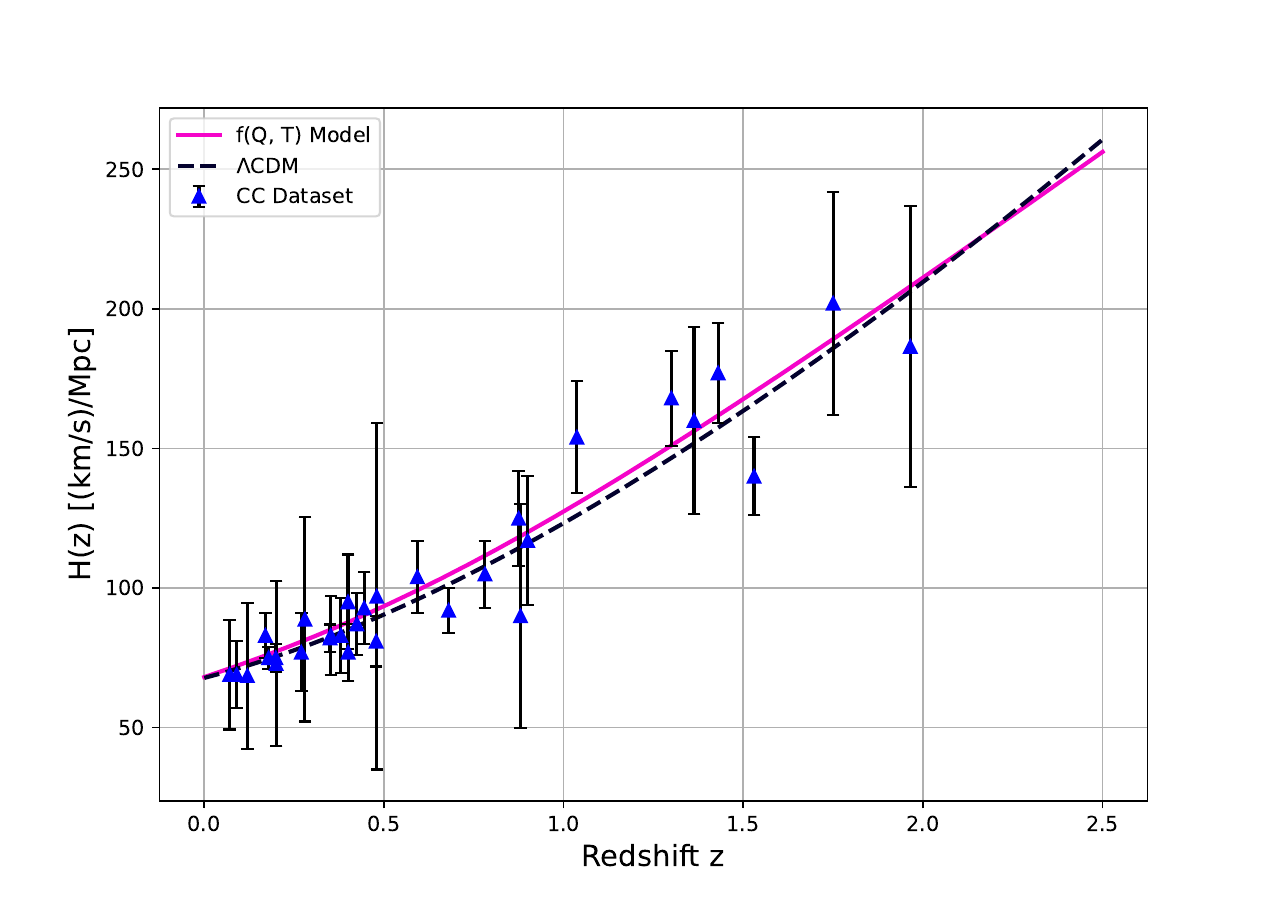}
\includegraphics[width=8.5cm,height=5cm]{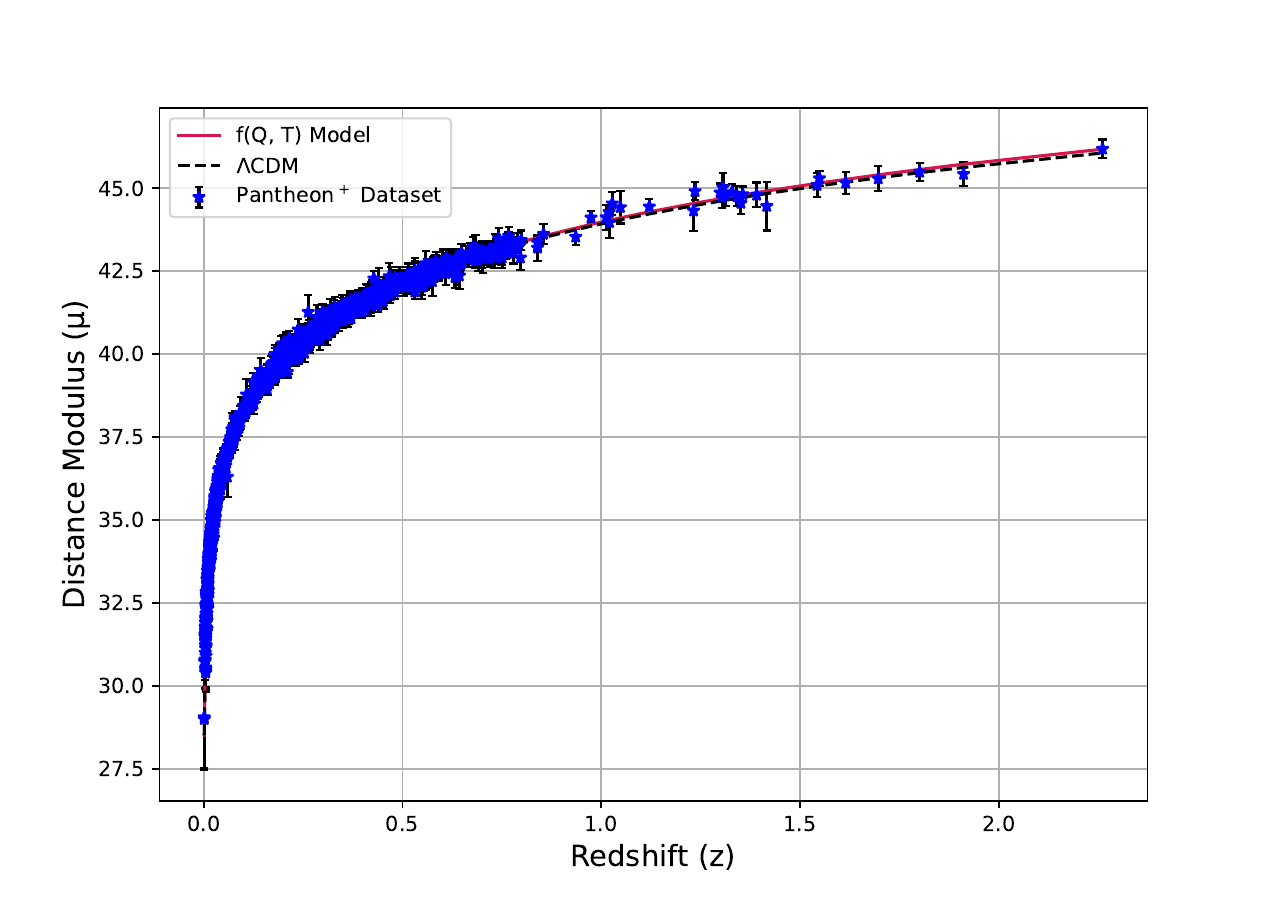}
{\caption{Evolution of Hubble parameter and distance modulus parameter for the model, $\Lambda$CDM and CC (upper Panel) and Pantheon$^{+}$ data sets (lower Panel).}}
\label{Fig2}
\end{figure}

\subsection{Results from Observational Data}
After performing MCMC analysis, we restricted the value of $ H_0 $, along with the model parameters $ \alpha $ and $ \beta $, using the CC and Pantheon$^+$ data sets. From this analysis, we obtained the minimum chi-square values ($ \chi^2_{\text{min}} $), reflecting how well the expansion history predicted by our model matches the observational data. For the CC data set, $ \chi^2_{\text{min}} $ is 16.51, while for the Pantheon$^+$ data set, it is 816.53, and for the combined CC$+$Pantheon$^+$ data set, it is 847.81. Additionally, we computed the AIC and BIC for the model. For the CC data set, the AIC and BIC values are 22.51 and 26.81, respectively; for the Pantheon$^+$ data set, they are 822.53 and 838.85; and for the combined CC $+$ Pantheon$^+$ data set, they are 859.81 and 880.12, respectively. The chi-square, AIC and BIC values computed for the $\Lambda$CDM model, as detailed in the \hyperref[Appendix_II]{Appendix-II}.

The analysis further revealed that the estimated present day Hubble parameter $ H_0 $ is $ 68.1 $ $\text{Km} \, \, \text{s}^{-1} \, \text{Mpc}^{-1}$ for the CC data set; $ 72.31 $ $\text{Km} \, \, \text{s}^{-1} \, \text{Mpc}^{-1}$ for the Pantheon$^+$ data set; and $ 72.75 $ $\text{Km} \, \, \text{s}^{-1} \, \text{Mpc}^{-1}$ for the combined data set. These values align with those reported by Di Valentino et al. \cite{DiValentino_2021_131}. The model parameters obtained from the MCMC analysis are summarized in Table \ref{tableA1}, while Fig. \ref{Fig1} presents contour plots showing the 1$\sigma$ and 2$\sigma$ confidence regions, along with comparisons of AIC, BIC, and $ \chi^2_{\text{min}} $ values against the $\Lambda$CDM model. Based on these comparisons, we conclude that our model closely mimics the $\Lambda$CDM model.

In Fig. \ref{Fig2} (upper panel), we plot the error bars for 32 data points from the CC data set, showing the behavior of the Hubble parameter for both the exponential model and the $\Lambda$CDM model. The lower panel displays a similar error bar plot for 1701 points from the Pantheon$^+$ data set.
\begin{table}
\renewcommand\arraystretch{1.5}
\centering % used for centering table
{
\begin{tabular}{|c|c|c|c|} % centered columns (3 columns)
\hline %inserts double horizontal lines
   \parbox[c][1.5cm]{2cm}{Cosmological Parameters
    }& \parbox[c][1.5cm]{1.8cm}{CC}  &  \parbox[c][1.5cm]{1.8cm}{Pantheon$^+$} & \parbox[c][1.5cm]{2.5cm}{CC$+$Pantheon$^+$}  \\ [0.3ex] % inserts table %heading% inserts table
%heading
% inserts single horizontal line
\hline\hline
$q_0$ & $-0.367$ &  $-0.396$ & $-0.353$ \\[0.5ex] % [1ex]
\hline %inserts single line
$w_0$ & $-0.578$ &  $-0.597$ & $-0.568$ \\
\hline
$r_0$ & $-0.281$ &  $-0.244$ & $-0.308$ \\[0.5ex] % [1ex]
\hline %inserts single line
$s_0$ & $0.493$  &  $0.463$ & $0.510$ \\[0.5ex] % [1ex]
\hline %inserts single line
\end{tabular}}
\caption{Cosmological parameter of $f(Q,T)$ model for CC, Pantheon$^+$ and combined data set.}\label{tableA2}% title of Table
\end{table}
\begin{figure}[H]
    \centering
    \includegraphics[width=8cm]{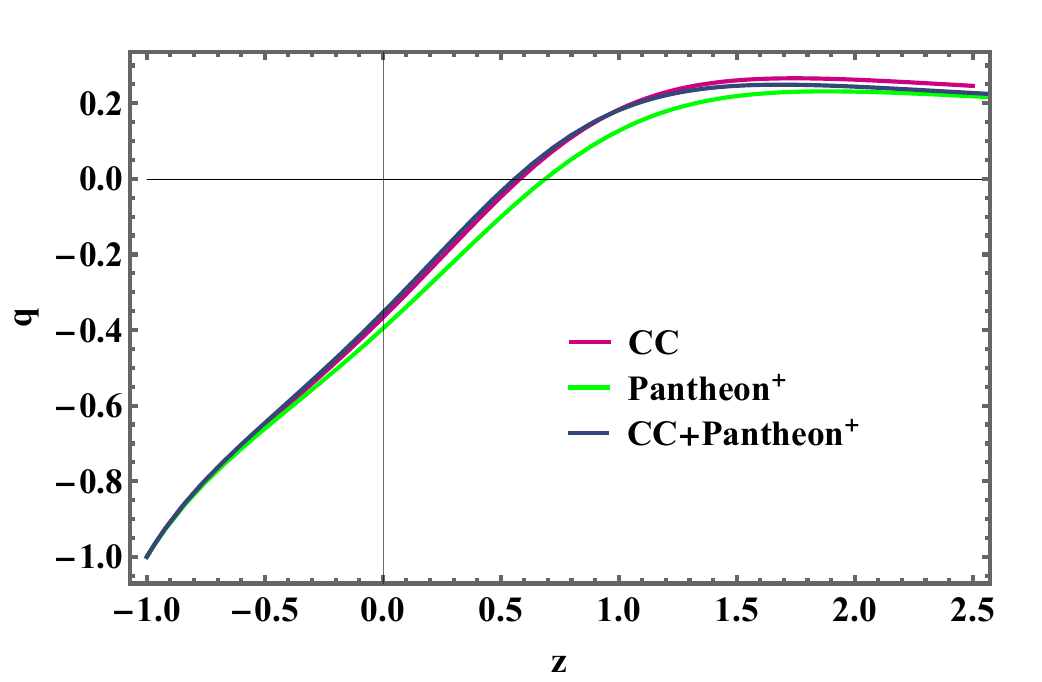}
{    \caption{Behavior of the deceleration parameter in redshift}}
    \label{Fig3} 
\end{figure}
  \begin{figure}[H]
    \centering
    \includegraphics[width=8cm]{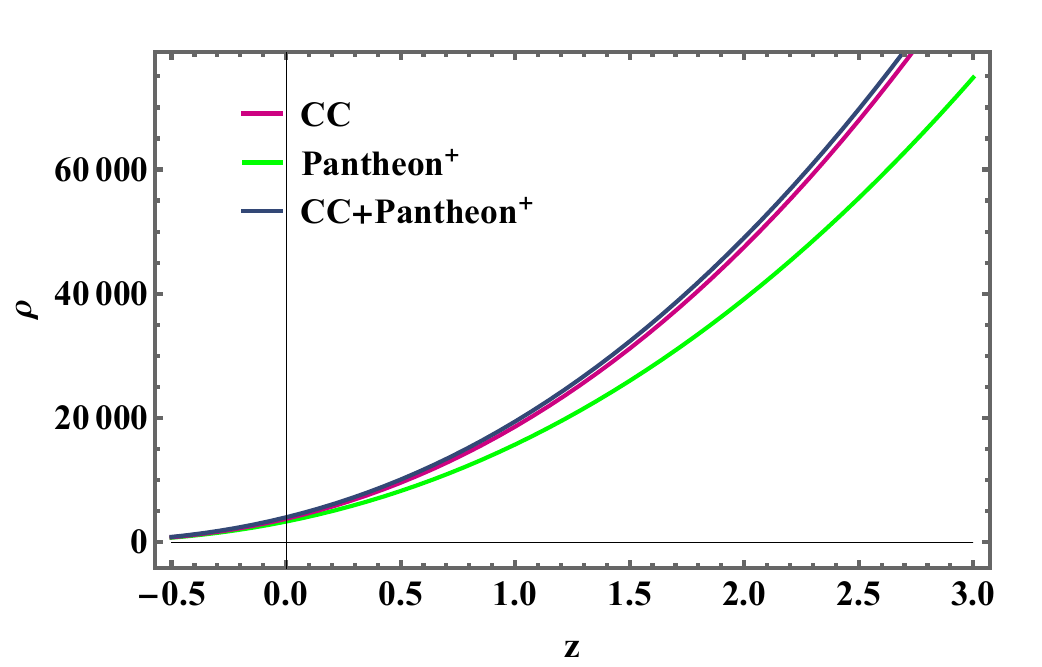}
  { \caption{Behavior of the matter density in redshift }}
    \label{Fig4} 
 \end{figure}
  \begin{figure}[H]
    \centering
    \includegraphics[width=8cm]{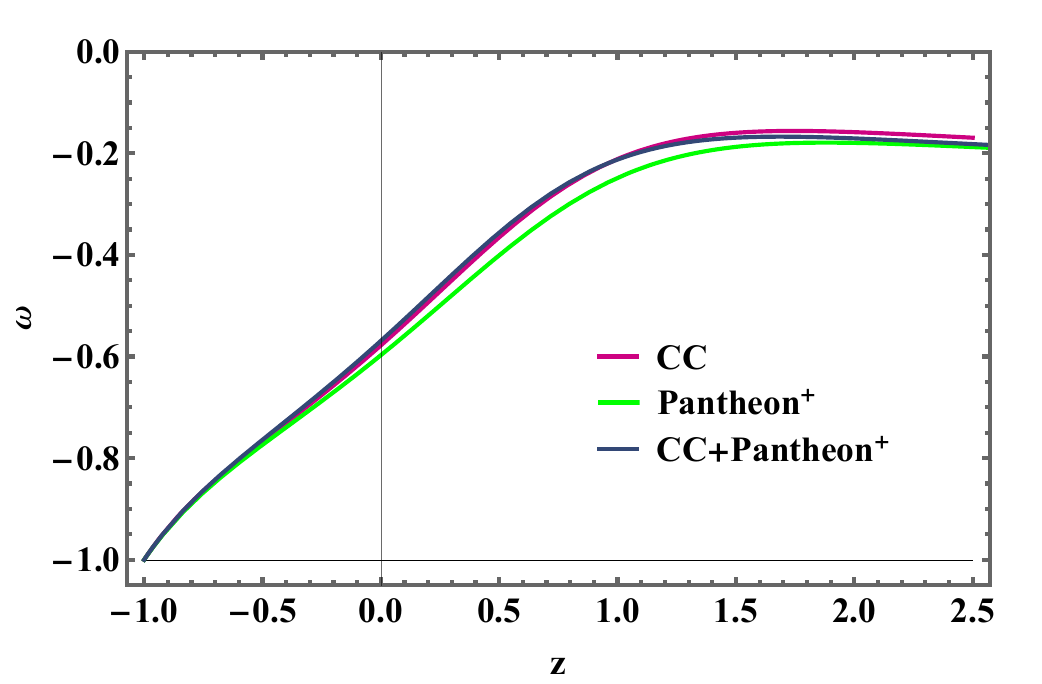}
    {\caption{Behavior of EoS parameter in redshift}}
    \label{Fig5} 
 \end{figure}
 
In addition to the analysis of the Hubble parameter, we examined other cosmological parameters to validate and further characterize the model. To demonstrate the current accelerated expansion of the Universe, we analyzed the deceleration parameter, which transitions from deceleration in the early Universe to acceleration in later times  [Fig. \ref{Fig3}]. The present values of the deceleration parameter are $ -0.367 $, $ -0.396 $, and $ -0.353 $, with transition redshifts $ z_{\text{tr}} = 0.581 $, $ z_{\text{tr}} = 0.688 $, and $ z_{\text{tr}} = 0.560 $ for the CC, Pantheon$^+$, and combined data sets, respectively. The matter density, which remains positive throughout the evolution of the Universe and approaches zero at late times [Fig. \ref{Fig4}]. The negative values of the deceleration parameter, approaching $-1$, along with the behavior of the equation of state (EoS) parameter, indicate ongoing acceleration, consistent with the predictions of the $\Lambda$CDM model [Fig. \ref{Fig5}].

\subsection{Statefinder Parameter}
The statefinder parameter is a cosmological diagnostic that aids in differentiating between various dark energy and modified gravity models by examining the dynamics of expansion of the Universe \cite{Sahni_2003_77}. It extends traditional parameters like the Hubble parameter and deceleration parameter by incorporating higher derivatives of the scale factor. Specifically, it introduces the dimensionless parameters $r$ and $s$, which provide a deeper understanding of the underlying mechanisms of cosmic acceleration. The jerk parameter $r$ is defined as
\begin{eqnarray}
r = \frac{\dddot{a}}{aH^3} \, ,
\end{eqnarray}
while the parameter $ s $ is given by
\begin{eqnarray}
s = \frac{r - 1}{3(q - \frac{1}{2})} \, .
\end{eqnarray}

In the standard $\Lambda$CDM model, these parameters are fixed at $r = 1$ and $s = 0$ \cite{Alam_2003}. As a result, $\Lambda$CDM serves as a reference point in the $r-s$ plane. Deviations from these values signify departures from the standard model, allowing for comparisons with alternative cosmological models such as quintessence, Chaplygin gas, or modified gravity frameworks.
  \begin{figure}[H]
    \centering
    \includegraphics[width=8cm]{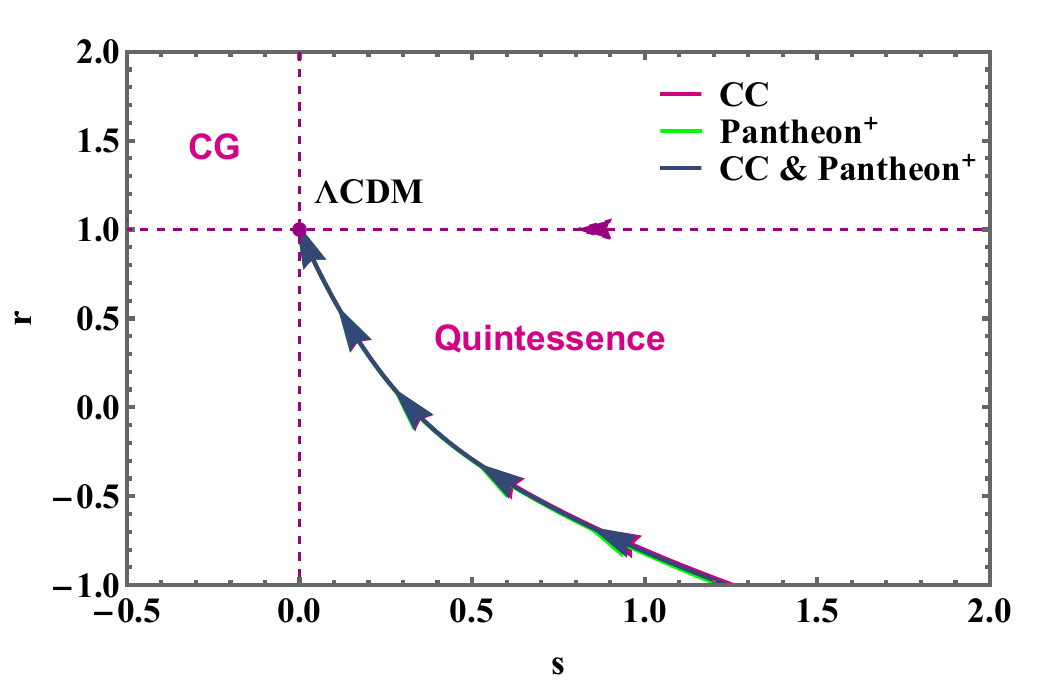}
    {\caption{Behavior of $r-s$ plot.}}
    \label{Fig6} 
 \end{figure}

The statefinder approach is particularly valuable as it takes into account higher-order expansions, making it more sensitive to the intricate physics driving cosmic acceleration. By plotting the evolution of $r$ and $s$ in the statefinder plane, each model traces a distinct trajectory, which enables a detailed assessment of its consistency with observational data. In Fig. \ref{Fig6}, we illustrate the behavior of the statefinder parameters for the $f(Q, T)$ gravity model, tracing the $r-s$ trajectory over a redshift range from $z = -1$ to $z = 3$. The path shows a transition from quintessence and ultimately converging to the $\Lambda$CDM model. The current values of $r$ and $s$ for different data sets are presented in Table \ref{tableA2}. This method offers a powerful framework for distinguishing between models that might produce similar results for lower-order parameters, like the Hubble parameter $H$ or the deceleration parameter $q$, but exhibit significantly different dynamic behaviors at higher orders.

\subsection{Age of the Universe}\label{Sec:AoU}
In the framework of $f(Q, T)$ gravity model, the age of the Universe is determined by evaluating the integral $ t_0 = \int_0^\infty \frac{dz}{H_0 (1+z)H(z)} $, which accounts for the cumulative contribution of the cosmic expansion rate throughout history. This calculation yields an age of approximately 13.831 Gyr for the CC data set, 13.898 Gyr for the Pantheon$^+$ data set, and 13.851 Gyr for the combined CC$+$Pantheon$^+$ data set, consistent with current observational data. These age estimates align well with the chronometric age of the star BD +17$^\circ$3248. The average age from various chronometric pairs suggests an age of $13.8 \pm 4$ Gyr for this star \cite{Cowan_2002}, in agreement within error limits with other chronometric age determinations for ultra metal-poor (UMP) and metal-poor Galactic halo stars. However, the relatively large uncertainties reflect the sensitivity of age estimates to both predicted and observed abundance ratios. Addressing these uncertainties necessitates additional stellar observations and advanced theoretical investigations. Ongoing efforts in this direction aim to improve the precision of age determinations for the oldest stars, thereby refining the estimated age of the Galaxy and imposing more stringent constraints on cosmological age estimates. This method effectively integrates the dynamic changes in the expansion rate, which are influenced by the dominance of different components over cosmic time.
   \begin{figure}[H]
    \centering
    \includegraphics[width=8cm]{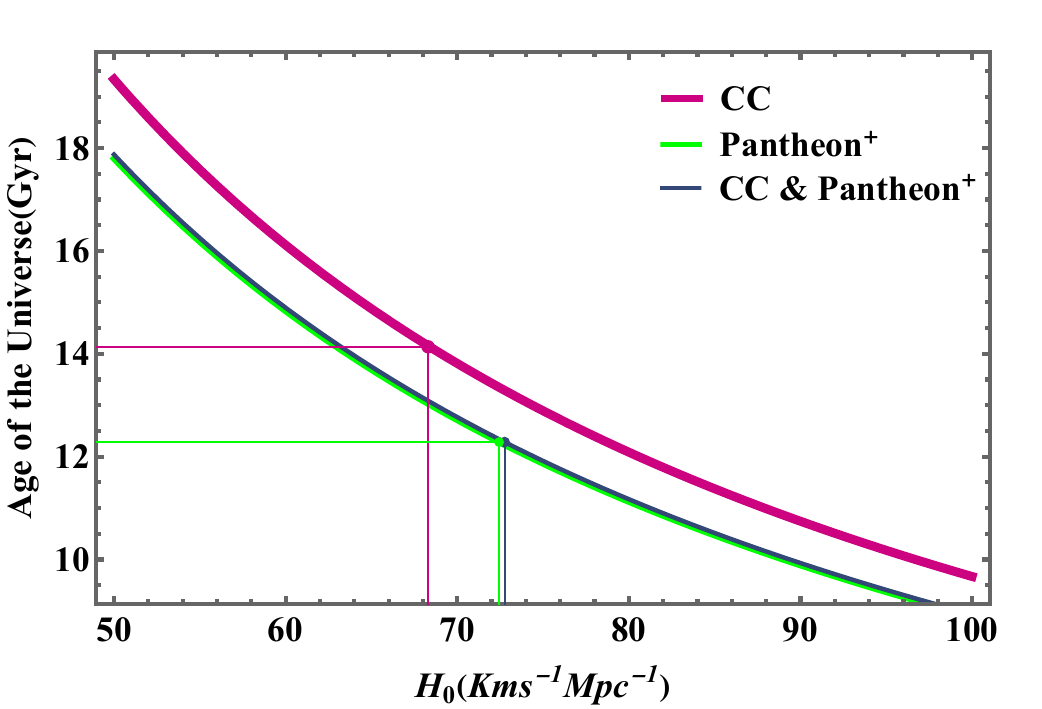}
    {\caption{Plot for the age of the Universe.}}
    \label{Fig7} 
 \end{figure}
 
\subsection{Energy Conditions}\label{Sec:EC} 
\begin{figure}[H]
    \centering
    \includegraphics[width=8cm]{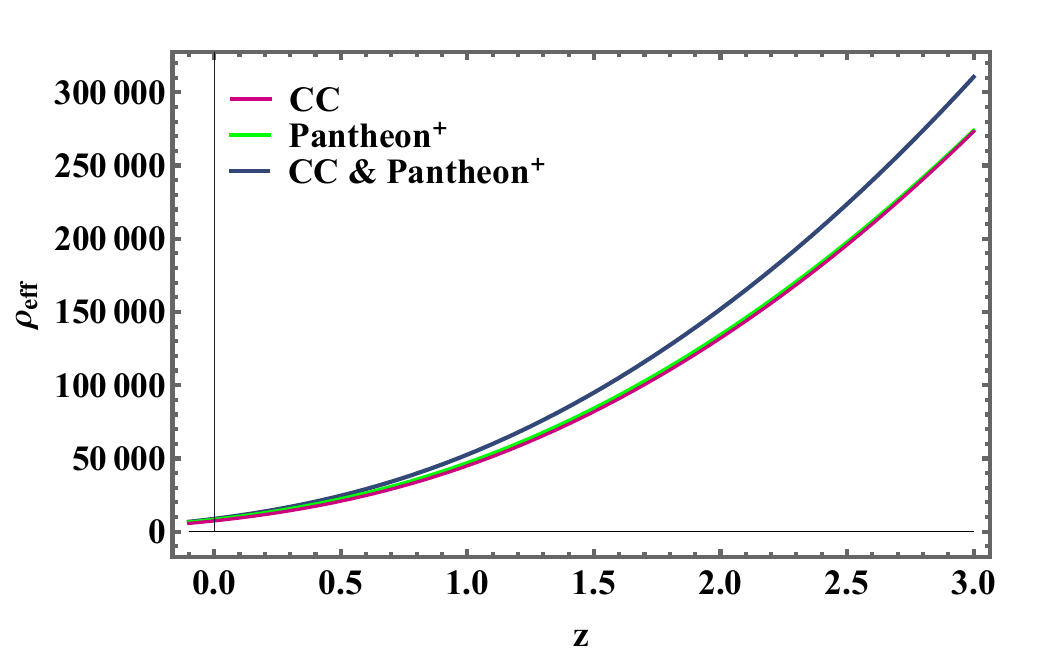}
   {\caption{Behavior of Effective Density}}
    \label{Fig8} 
\end{figure}

In the $ f(Q, T) $ gravity model, we examine the behavior of effective pressure $ p_{\text{eff}} $ and effective density $ \rho_{\text{eff}} $ to test key energy conditions. Fig. \ref{Fig8} illustrates the behavior of the effective density parameter, which remains positive throughout the entire cosmic evolution, consistent with the expected dynamics of the expansion of the Universe. The null energy condition (NEC), $ \rho_{\text{eff}} + p_{\text{eff}} \geq 0 $, is satisfied throughout cosmic evolution, ensuring physically consistent behavior. Similarly, the weak energy condition (WEC), requiring $ \rho_{\text{eff}} \geq 0 $, and the dominant energy condition (DEC), $ \rho_{\text{eff}} - p_{\text{eff}} \geq 0 $, hold during the evolution, ensuring non-negative energy density and causal propagation.
{
\begin{eqnarray}
         \rho_\text{eff} = \frac{3~\beta   }{2}\rho(z)+3~\alpha~ {H_0}^2 e^{\frac{\alpha  ~{H_0}^2}{H(z)^2}}+\rho(z),\nonumber\\
\end{eqnarray}
\begin{eqnarray}
         p_\text{eff}&=&-\frac{\beta ~ \rho(z) }{2}-\frac{\alpha  {H_0}^2 e^{\frac{\alpha  {H_0}^2}{H(z)^2}} \left(4 \alpha  {H_0}^2 (z+1) H'(z)+3 H(z)^3\right)}{H(z)^3}.\nonumber \\
\end{eqnarray}}
\begin{widetext}
\begin{figure}[H]
    \centering
    \includegraphics[width=18cm]{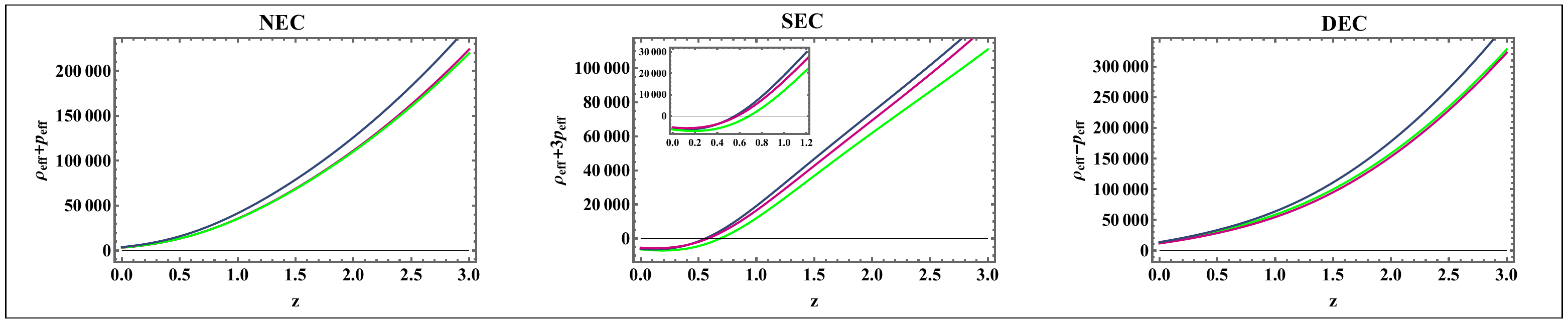}
   {\caption{Behavior of energy condition for different data sets.}}
    \label{Fig9} 
 \end{figure}
\end{widetext}

However, the SEC, $ \rho_{\text{eff}} + 3p_{\text{eff}} \geq 0 $, is violated at the late time shown in Fig. \ref{Fig9} (Evolution equation of energy conditions given in \hyperref[Appendix_I]{Appendix-I}). This violation is consistent with the observed accelerated expansion of the Universe, as the SEC traditionally implies decelerated expansion. The violation of the SEC aligns with observational evidence, confirming the ability of the $ f(Q, T) $ gravity model to describe the transition to late-time cosmic acceleration effectively.

\section{Conclusion}\label{Sec:4} 

In this paper, we have investigated an exponential model within the framework of $f(Q, T)$ gravity, an extension of symmetric teleparallel gravity. This model introduces a new class of theories where the nonmetricity scalar $Q$ is coupled non-minimally to the trace of the energy-momentum tensor $T$. To ensure a consistent and general formulation of the theory, we adopted the covariant formulation of $f(Q, T)$ gravity, which allows the field equations to be expressed in a coordinate-independent manner. Utilizing the metric-affine formalism, we conducted our analysis, where the coupling between $Q$ and $T$ leads to the non-conservation of the energy-momentum tensor. We derived the expansion function (Hubble parameter) as a first-order ODE and obtained the matter density from the modified Friedmann equations. The model was studied using CC, Pantheon$^+$, and combined CC$+$Pantheon$^+$ datasets. Using MCMC analysis, we constrained the model parameters $\alpha$, $\beta$, and $H_0$ with the help of observational data. 

For validation, we compared our results with the standard $\Lambda$CDM model, which is currently the most widely accepted cosmological model. In this study, we analyzed the Hubble parameter and plotted the distance modulus. We also examined cosmographic parameters, particularly the deceleration parameter, which indicates the transition of the Universe from deceleration to acceleration, approaching a value of $-1$ at late times. The negative value of the EoS parameter confirms that the Universe is currently in a dark energy-dominated phase, represented by this modified gravity model.

Furthermore, we explore the statefinder diagnostic and presented the behavior of the $r-s$ plot for the $f(Q, T)$ model over a redshift range from $-1$ to 3. The trajectory shows a transition from quintessence phase before converging to the $ \Lambda $CDM model. By integrating the inverse Hubble parameter as a function of redshift, we calculated the current age of the Universe, obtaining $t_0 = 13.831$ Gyr for the CC data set, $ t_0 = 13.898$ Gyr for Pantheon$^+$, and $t_0 = 13.851$ Gyr for the combined data set.

In addition, the violation of the SEC, $ \rho_{\text{eff}} + 3p_{\text{eff}} \geq 0 $, aligns with the observed acceleration of the Universe. This violation is consistent with observational evidence, demonstrating the ability of the $ f(Q, T) $ gravity model to effectively describe the transition to late-time cosmic acceleration. The inclusion of the covariant formulation in this work not only strengthened the theoretical foundation of the model but also enabled a more precise interpretation of the cosmological implications within the context of modified gravity.

\section*{Acknowledgement} RB acknowledges the financial support provided by the University Grants Commission (UGC) through Junior Research Fellowship UGC-Ref. No.: 211610028858 to carry out the research work. SVL would like to express gratitude for the financial support provided by the University Grants Commission (UGC) through the Senior Research Fellowship (UGC Reference No.: 191620116597) to carry out the research work.  BM acknowledges the support of IUCAA, Pune (India), through the visiting associateship program.
\begin{widetext}
 \section*{Appendix-I}\label{Appendix_I}
Energy conditions in terms of the Hubble parameter in redshift.
{
\begin{eqnarray}
    \rho_{eff}+p_{eff}&=&(\beta +1)~ \rho(z) -\frac{4~ \alpha ^2~ H_0^4 ~(z+1)~ H'(z)~ e^{\frac{\alpha~  H_0^2}{H(z)^2}}}{H(z)^3}\, ,\\
    \rho_{eff}+3p_{eff}&=&-\frac{12 \alpha ^2 ~H_0^4~ (z+1)~ H'(z) ~e^{\frac{\alpha~  H_0^2}{H(z)^2}}}{H(z)^3}-6 ~\alpha~  H_0^2 e^{\frac{\alpha~  H_0^2}{H(z)^2}}+\rho(z) \, ,\\
    \rho_{eff}-p_{eff}&=&2 ~\beta~  \rho(z) +\frac{4 \alpha ^2~ H_0^4 ~(z+1)~ H'(z) e^{\frac{\alpha~  H_0^2}{H(z)^2}}}{H(z)^3}+6~ \alpha~  H_0^2~ e^{\frac{\alpha~  H_0^2}{H(z)^2}}+\rho(z)\, ,
\end{eqnarray}}
\end{widetext}
\section*{Appendix-II}\label{Appendix_II}
\begin{figure}[H]
    \centering
    \includegraphics[width=76mm]{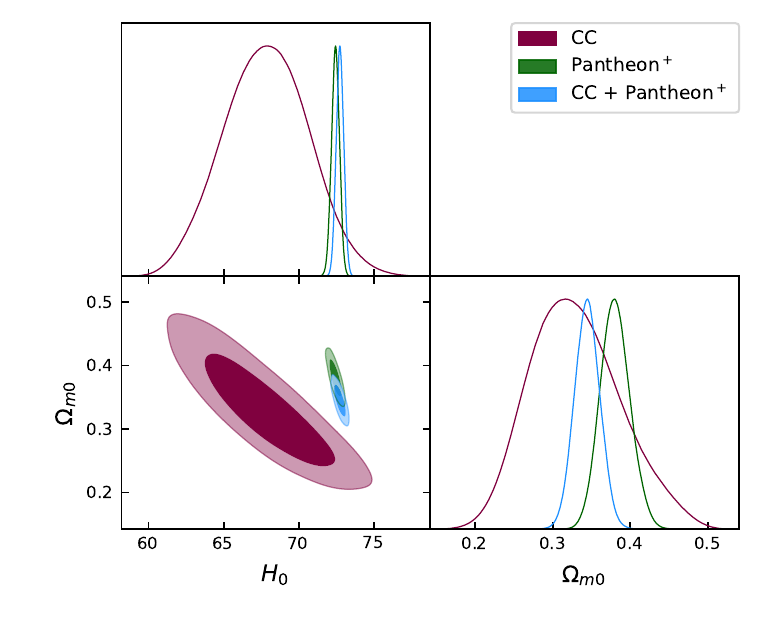}
    \caption{$\Lambda$CDM contour plot}
    \label{Fig10} 
\end{figure}

In our comparison of the exponential $ f(Q, T) $ gravity model with the CC, Pantheon$^+$, and combined CC$+$Pantheon$^+$ data sets, we also present the corresponding MCMC results for the $\Lambda$CDM model for reference. The $\Lambda$CDM results are provided for transparency. Fig. \ref{Fig10} illustrates the MCMC posterior distributions and confidence regions for different priors applied to each data set. The analysis shows rapid convergence across all priors and data sets. From this, we determine the values of the Hubble constant $H_0 $ and the matter density parameter $\Omega_{m,0} $. For the CC data set, we find $H_0 = 67.80 \pm 2.8$ $\text{Km} \, \, \text{s}^{-1} \, \text{Mpc}^{-1}$ and $\Omega_{m,0} = 0.329^{+0.050}_{-0.066} $. For the Pantheon$^+$ data set, the values are $H_0 = 72.44 \pm 0.26$ $\text{Km} \, \, \text{s}^{-1} \, \text{Mpc}^{-1}$ and $\Omega_{m,0} = 0.38^{+0.018}_{-0.020} $, while for the combined CC$+$Pantheon$^+$ data set, we obtain $H_0 = 72.74 \pm 0.25$ $\text{Km} \, \, \text{s}^{-1} \, \text{Mpc}^{-1}$ and $\Omega_{m,0} = 0.344 \pm 0.016 $.
 
The minimum chi-square values ($\chi^2_{\text{min}} $) are 14.54 for the CC data set; 812.61 for the Pantheon$^+$ data set; and 846.63 for the combined CC$+$Pantheon$^+$ data set. The AIC and BIC values for the data sets are as follows: for the CC data set, AIC = 18.54 and BIC = 21.41; for the Pantheon$^+$ data set, AIC = 816.61 and BIC = 827.49; and for the combined data set, AIC = 854.63 and BIC = 868.37. These metrics provide important insights into the performance of the model and the selection criteria based on the data sets used.

\bibliographystyle{utphys}
\bibliography{references}

\end{document}